# A PREFIXED-ITEMSET-BASED IMPROVEMENT FOR APRIORI ALGORITHM

Yu Shoujian[1], Zhou Yiyang[2]

[1]College of computer science and technology, Donghua University, Shanghai, 201600, China
`jackyysj@dhu.edu.cn`
[2] College of computer science and technology, Donghua University, Shanghai, 201600, China
`yiyang0203@foxmail.com`

## ABSTRACT

*Association rules is a very important part of data mining. It is used to find the interesting patterns from transaction databases. Apriori algorithm is one of the most classical algorithms of association rules, but it has the bottleneck in efficiency. In this article, we proposed a prefixed-itemset-based data structure for candidate itemset generation, with the help of the structure we managed to improve the efficiency of the classical Apriori algorithm.*

## KEYWORDS

*Data mining, association rules, Apriori algorithm, prefixed-itemset, hash map*

## 1. INTRODUCTION

With the rapid development of computer technology in various sectors, the data generated by different industries are becoming more and more, but how to get valuable information from the big data has become a new problem. Data mining, that is data knowledge discovery, came into being in this backdrop. Data mining is to excavate the implied, unknown, interesting knowledge and rules from a large number of data [1]. Association rules is an important part of data mining, it was first put forward by R.Agrawal, mainly to solve the customer transaction association rules between sets of items in the transaction library [2]. In the following year, R.Agrawal proposed the most classical algorithm to calculate association rules, that is Apriori algorithm [3], which is to infer the (k+1) – itemsets by the k- itemsets.

However, due to the computing bottleneck of Apriori algorithm when calculating the candidate set, in recent years there have been many improved algorithms of the traditional Apriori algorithm from different aspects. Chun-Sheng Z proposed an improved Apriori algorithm based on classification [4]. Jia Y improves the algorithm from the aspect of transaction database partitioning and dynamic itemset planning [5]. Shuangyue L proposed an improved algorithm based on the matrix of database to enhance the efficiency of calculating [6]. Wang P proposed an optimization method to reduce the search times of the transaction library to improve the efficiency [7]. Vaithiyanathan V uses the method of compressing the transactions of the similar interests in the database to improve the efficiency of the algorithm [8]. Lin X implements Apriori algorithm based on Map Reduce to improve the candidate sets of large amounts of data generation efficiency [9]. Zhang first analyze the characteristic of the data, that is medical data, and then combine the characteristics of the data to improved Apriori algorithm [10]. Wu Huan proposed an improved algorithm IAA, which adopts a new count-based method to prune candidate itemsets and uses generation record to reduce total data scan amount [11]. Wang Yuan proposes an improved item constrain association rules mining algorithm, which improves traditional algorithm in two aspects: trimming frequent itemsets and calculating candidate itemsets [12]. Lin Ming-Yen proposes three algorithms, named SPC, FPC, and DPC, to investigate effective implementations of the Apriori algorithm in the MapReduce framework [13]. Chai Sheng proposes a novel algorithm so called Reduced Apriori Algorithm with Tag (RAAT), which reduces one redundant pruning operations of $C2$ [14].

This article will be focus on the two concrete steps of classical Apriori algorithm, namely connecting step and the pruning step, using a new prefix-itemset-based storage, combining the fast lookup feature of hash tables to improve the efficiency. This paper will first describe the classical Apriori algorithm and its shortcomings, then specifically describe the improvements, and finally introduce the comparisons of efficiency of classical Apriori algorithm and improve Apriori algorithm on specific data sets.

## 2. APRIORI ALGORITHM

### 2.1. Apriori algorithm introduction

Apriori algorithm is a classical algorithm for frequent itemset mining association rules, the basic idea of the algorithm is to use an iterative approach layer by layer to find the frequent. The algorithm will first obtain k-itemsets, and then use the k- itemsets to explore (k+1)-itemsets. First, let's introduce the priori knowledge of frequent itemsets, which is, any subset of a frequent itemset is also a frequent itemset. Apriori algorithm uses the prior knowledge of frequent itemsets, first to find the collection of frequent 1-itemsets, denoted $L_1$. Then use the 2-itemsets of $L_1$ to get $L_2$, and then $L_3$, and so on, until you cannot find the frequent k-itemsets. Apriori algorithm mainly consists of the following three steps:

(1) Connecting step: connecting k- frequent itemsets to generate (k+1)-candidate sets, denoted by $C_{k+1}$. The connect condition of the connecting step is that the two k-itemsets have the same first (k-1) items and different k-th items. Denote $l_i[j]$ is j-th item of $l_i$, the condition is:

$$l_1[1] = l_2[1] \wedge l_1[2] = l_2[2] \wedge \ldots \ldots \wedge l_1[k-1] = l_2[k-1] \wedge l_1[k] \neq l_2[k]$$

In which $l_1$ and $l_2$ are k-item subset of the set collection $L_k$, $l_1[k] \neq l_2[k]$ is to ensure not to generate duplicate k- itemsets. Itemsets generated by the $l_1$ and $l_2$ connection as follows:

$$\{l_1[1], l_1[2], l_1[3], \ldots \ldots, l_1[k], l_2[k]\}$$

(2) Pruning step: To pick out the true frequent itemsets $L_{k+1}$ from the candidate set $C_{k+1}$. Because the candidate set $C_{k+1}$ is the superset of the true frequent itemsets $L_{k+1}$. According to the nature of Apriori: any subsets of frequent set must also be frequent, that is any (k-1)- items subsets of k-items must also be frequent. With this property we can find out if the k- items subsets of $C_{k+1}$ are in $L_k$, if not, then remove the candidate (k + 1) - itemset is removed from the $C_{k+1}$.

(3) Counting step: scanning the database, accumulate the number of candidates appearing in the database. If the appear times of a candidate is less than the given minimum support threshold, the candidate itemset will be removed.

### 2.2. Shortage of Apriori algorithm

Apriori is one of the most classical algorithms for mining association rules, but it also has the shortage of low efficiency. The time Apriori algorithm consumes lies mainly in the following three aspects:

(1) In connection step, when connects k-itemsets to generate (k+1)-itemsets, it compares too many times to determine if the itemsets meets the connection conditions. When $L_k$ has m k- itemsets, the time complexity of the connection step is $O(k*m^2)$.

(2) In the pruning step, when determine if a subset of candidate set $C_{k+1}$ is in the frequent set $L_k$, the best situation is to simply scan once to get the result, while the worst-case is that is needs to scan k times to find that the k-th subset of $C_{k+1}$ is not in the $L_k$. So the average times need to scan and compare the $L_k$ is $|C_{k+1}| * |L_k| * k / 2$.

(3) In counting step, when accumulate the support times of itemsets in $C_{k+1}$, we need to scan the database for $|C_{k+1}|$ times.

Taking into account these three aspects of time-consuming steps of classical Apriori algorithm, this article presents an improved Apriori algorithm based on prefix-itemset.

## 3. IMPROVED APRIORI ALGORITHM

### 3.1. Improved Apriori algorithm

In 1.2 we have analyzed the shortcomings of classical Apriori algorithm, so its improvements also focus on the three steps mentioned in 1.2. Since the records are already sorted by the dictionary, therefore the candidate set generated by Apriori algorithm is ordered.

（1）Prefixed-itemset-based storage

In the improved algorithm we proposed a new method to store the itemsets. For each itemset in $L_k$, we use a structure similar to Map <key, value> to store them, in which we save the forward (k-1)- item content as the key while the last item content as the value. After having all the itemsets saved in the new format, we group all the itemsets with the same key and store the union of their values as the new value.

For example: the database is shown in Table 1 and the minimum support is 2.

Table 1 database

| TID | Itemset |
|---|---|
| T1 | A,B,E |
| T2 | B,D |
| T3 | B,C |
| T4 | A,B,D |
| T5 | A,C |
| T6 | B,C |
| T7 | A,C |
| T8 | A,B,C |
| T9 | A,B,C,E |

The traditional Apriori algorithm will scan the database to obtain the times each item appears in the database, to form the 1- itemsets, and then to generate the 2-itemsets that meets the minimum support, that is 2. Here is the content generated by the classical Apriori algorithm.

Table 2 classical Apriori algorithm

| 1-itemset | | 2- itemset | |
|---|---|---|---|
| Item | Count | Item | Count |
| A | 6 | AB | 4 |
| B | 7 | AC | 4 |
| C | 6 | AE | 2 |
| D | 2 | BC | 4 |
| E | 2 | BD | 2 |
| | | BE | 2 |

While Table 3 shows how we store the itemsets with the prefix-itemset-based storage.

Table 3 prefix-itemset-based storage

| | Prefixed-key | Value |
|---|---|---|
| 1-itemset | NULL | {A, B, C, D, E} |
| 2-itemset | A | {B, C, E} |
| | B | {C, D, E} |

As shown in Table 3, 1- itemset has only one item, so the key of 1- itemset is NULL. Besides, we can infer the length of the itemset from the length of the key because the length of the value of the key stores all the items in the itemset but the last item.

(2) Prefixed-itemset-based connecting step

After the establishing of prefix-itemset-based storage, when we have to generate (k+1)-itemset by connecting the two k- itemsets, we can simply combine two different items in the value, and then generate new itemset with the key. For example, when connecting the 2-itemset with the prefix-key of A in Table 3, we can generate the 3- itemset by combine the value and get the result as {{B, C},{B, D},{C, D}}.

(3) Prefixed-itemset-based pruning step

In chapter 1.1 we know that (k+1)- itemsets are generate from two k- itemsets, and if any k-itemset subset of the (k+1)-itemset does not exist in $L_k$, then we have to remove the (k+1)- itemset from $C_{k+1}$.

**Theorem:** If we generate a (k+1)-itemset by connecting two k-itemsets, $l_1$ and $l_2$, and one k- itemset of all the k-itemset subset does not exist in $L_k$, then the k-itemset subset must contains both $l_1[k]$ and $l_2[k]$.

**Prove**: Assume $l_1$ and $l_2$ are both k-itemset, and the (k+1)-itemset generated by connecting $l_1$ and $l_2$ is $\{l_1[1], l_1[2], l_1[3], ......, l_1[k], l_2[k]\}$. If the k- itemset dose not contains both $l_1[k]$ and $l_2[k]$, then the possible options are $\{l_1[1], l_1[2], l_1[3], ......, l_1[k]\}$ and $\{l_1[1], l_1[2], l_1[3], ......, l_2[k]\}$, that is $l_1$ and $l_2$, and both $l_1$ and $l_2$ come from $L_k$, so if the k-itemset does not belong to $L_k$, then it must contains both $l_1[k]$ and $l_2[k]$.

So in prefixed-itemset-based pruning step, we can simply consider the subset of (k+1)-itemset which contains both the last two items. With the example from Table 3, we can get the result as follow.

Table 4 pruning step

| Subset of 3-itemset | If belong to $L_2$ |
|---|---|
| B,C | yes |
| B,E | yes |
| C,E | no |
| C,D | no |
| C,E | no |
| D,E | no |

As shown in Table 4, only {{B,C},{B,E}} are possible 2-itemset subsets, plus the corresponding prefix-key, that is {{A,B,C},{A,B,E}}, namely the candidate set $C_3$.

After the pruning step, we have to scan the database to accumulate the times the itemset appears. After accumulating the times of itemsets after pruning step, we can find that both {A,B,C} and {A,B,E} meet the minimum support, and then we add them to the prefix-itemset-based storage as follows.

Table 5 3-itemset storage

| | prefix-key | Value |
|---|---|---|
| 3-itemset | A,B | {C,E} |

### 3.2. Algorithm

The algorithm is described as follow:

Input: Database D, minimum support min_sup
Output: frequent itemsetsL
1) $L_1$=1-itemset of D
2) Map<String[],String[]> map;
3) Import $L_1$ to map, set the key as null, value as the union of items in $L_1$
4) for(k=2;$L_{k-1}$ ≠ φ;k++){
5) $C_k$=pre_apriori_gen(map,k-2);
6) count the appear times of every itemset of $C_k$, $L_k$={c∈$C_k$|c.count>min_sup}
7) }
8) Return $L_k$;
procedurepre_apriori_gen(map:Map<String[],String[]>;k:int)
1) for each key in map{
2) if(key.length()==k){
3) c:=key plus two items from value
4) if(map.containsKey(c[0:k])){
5) If( any (k+1)- itemset subset belong to (key,value)){
6) put c into $C_k$
7) }
8) }
9) else continue;
10) }
11) Return $C_k$

# 4. EXPERIMENT AND RESULTS

The data of the experimental is a total of 120000 patients with diabetes clinical prescription data in Ruijin Hospital, the data records the prescription drug number per user per visit. This experimental machine is configured to Core i5 2.7GHz 8GB processor, 1866MHz LPDDR3 Intel memory.

This experiment compares the classical Apriori algorithm and the prefixed-itemset-based algorithm from two aspects, one is to compare the operation efficiency with fixed total tests and variable minimum support, the other is to compare the operation efficiency with fixed minimum support and variable total tests.

The result of the first experiment is as follows.

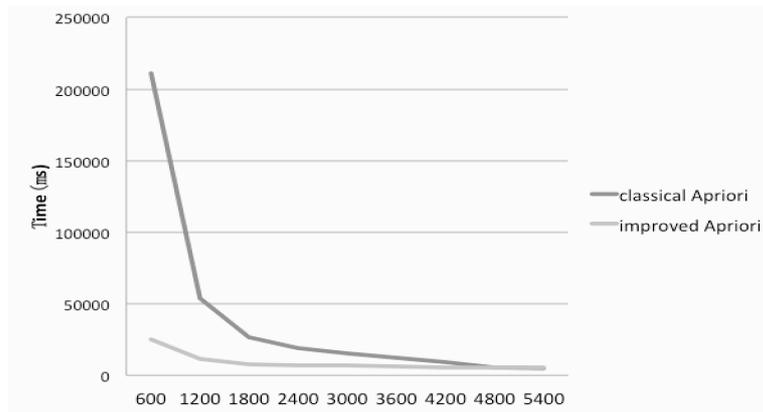

Figure 1. Time consuming with variable min_sup

The picture above shows the time consuming of the two algorithms when given fixed total test and variable min_sup. We can infer that the less min_sup is, the more operation efficiency the improved algorithm improves. And when the min_sup increases to a certain point, the classical Apriori algorithm and the improved algorithm are of the same efficiency.

Table 6 improvements under variable min_sup

| Min_sup (total 12w) | Classical Apriori Time(ms) | Improved Apriori Time(ms) | Improvement (%) |
|---|---|---|---|
| 600 | 210696 | 25192 | 81.44% |
| 1200 | 53822 | 11648 | 68.63% |
| 1800 | 26317 | 7614 | 51.88% |
| 2400 | 19359 | 7127 | 38.99% |
| 3000 | 15508 | 6753 | 56.45% |
| 3600 | 12393 | 5842 | 52.86% |
| 4200 | 9017 | 5424 | 39.85% |
| 4800 | 5705 | 5175 | 9.29% |
| 5400 | 4868 | 5161 | -6.02% |

Table 6 shows the specific operation time of the classical Apriori algorithm and the improved Apriori algorithm and the comparison between them.

The results of the second experiment are as follows.

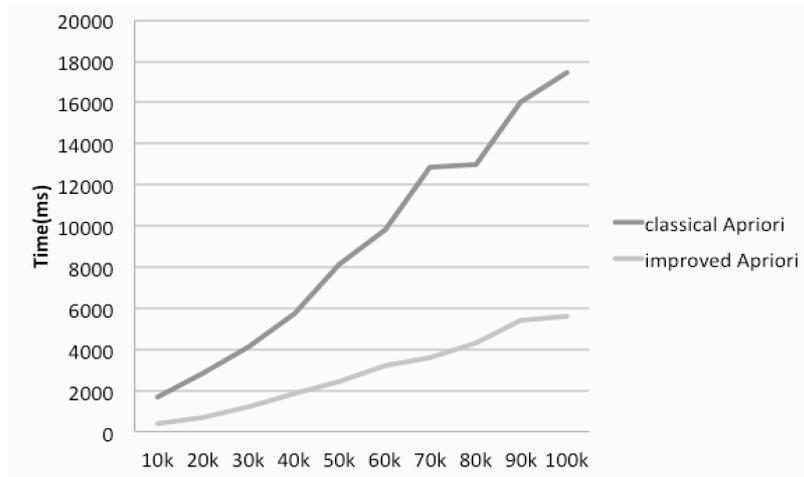

Figure 2. Time consuming with variable total test

The picture above shows the time consuming of the two algorithms when given fixed min_sup and variable total test. And we can tell that when the total test becomes larger, the improvements become more obvious.

Table 7 improvements under variable total test

| Total tests (min_sup% =2%) | Classical Apriori Time(ms) | Improved Apriori Time(ms) | Improvement (%) |
|---|---|---|---|
| 10k | 1686 | 390 | 76.87% |
| 20k | 2839 | 695 | 75.52% |
| 30k | 4088 | 1229 | 69.94% |
| 40k | 5729 | 1846 | 67.78% |
| 50k | 8141 | 2409 | 70.41% |
| 60k | 9833 | 3197 | 67.49% |
| 70k | 12848 | 3630 | 71.75% |
| 80k | 13004 | 4339 | 66.63% |
| 90k | 16007 | 5442 | 66.00% |
| 100k | 17438 | 5588 | 67.96% |

Table 7 shows the specific operation time of the two algorithm and we can learn from the table that when the min_sup is fixed to 2% of the total test, the improvement rate is about 70%.

Experiments have shown that the prefix-itemset-based Apriori algorithm is effective and feasible.

## 4. SUMMARY

In this paper, we described the Apriori algorithm specifically, and pointed out some limitations of the classical Apriori algorithm during the two steps of the algorithm, namely the connection and the paper cutting steps, and proposed the method of prefixed-itemset-based data storage and the improvements based on it. With the help of prefixed-itemset-based data storage, we managed to finish the connecting step and the pruning step of the Apriori algorithm much faster, besides we can store the candidate itemsets with smaller storage space. Finally, we compare the efficiency of classical Apriori algorithm and improve Apriori algorithm on the aspect of support degree and the total number, and the experimental results on both aspects proved the feasibility of the prefixed-itemset-based algorithm.

**Authors**

Yu Shoujian, the vice professor, main research direction: Web services, enterprise application integration, database and data warehouse;

Zhou Yiyang, Zhou Yiyang, master, the main research direction: data mining, machine learning